\documentclass[aps,prl,superscriptaddress,twocolumn,10pt,floatfix,lineno]{revtex4-1}
\usepackage{hyperref}

\usepackage{spfamilies}

\newcommand{\ifproofpre}[2]{#1}

\begin{document}

\title{Emergent $\grpsptr$ dynamical symmetry in the nuclear many-body system from an \emph{ab initio} description}

\author{Anna E. McCoy}
\affiliation{TRIUMF, Vancouver, British Columbia V6T 2A3, Canada}
\affiliation{Department of Physics, University of Notre Dame,
Notre Dame, Indiana 46556, USA}

\author{Mark A. Caprio}
\affiliation{Department of Physics, University of Notre Dame, Notre Dame, Indiana 46556, USA}

\author{Tom\'{a}\v{s} Dytrych}
\affiliation{Nuclear Physics Institute, Academy of Sciences of the Czech Republic, 250\,68 \v{R}e\v{z}, Czech Republic}
\affiliation{Department of Physics and Astronomy, Louisiana State University, Louisiana 70803, USA}

\author{Patrick J. Fasano}
\affiliation{Department of Physics, University of Notre Dame, Notre Dame, Indiana 46556, USA}

\date{\today}

\begin{abstract}
\textit{Ab initio} nuclear theory provides not only a microscopic framework for
quantitative description of the nuclear many-body system, but also a foundation
for deeper understanding of emergent collective correlations. A symplectic
$\grpsptr\supset\grpu{3}$ dynamical symmetry is identified in \emph{ab initio} predictions, from a
no-core configuration interaction approach, and found to provide a qualitative
understanding of the spectrum of $\isotope[7]{Be}$. Low-lying states form an Elliott $\grpsu{3}$
spectrum, while an $\grpsptr$ excitation gives rise to an excited rotational band
with strong quadrupole connections to the ground state band.

\end{abstract}


\maketitle



The nucleus is a complex many-body system, which nonetheless exhibits simple
patterns indicative of emergent collective
degrees of freedom~\cite{Ring1980,Talmi1993,Casten2000,RoweWood2010}.
\textit{Ab initio} nuclear theory now provides accurate quantitative predictions for
observables in light
nuclei~\cite{prl-84-2000-5728-Navratil,prc-76-2007-044305-Hagen,prc-79-2009-044606-Quaglioni,prc-86-2012-034321-Bacca,prl-111-2013-252501-Dytrych,prc-87-2013-034326-Baroni,rmp-87-2015-1067-Carlson,prc-93-2016-051301-Stroberg,prc-99-2019-064320-Henderson,npa-738-2004-357-Neff}.
Signatures of collective phenomena including
clustering~\cite{prc-70-2004-054325-Pieper,npa-738-2004-357-Neff,jpcs-402-2012-012031-Maris,fbs-54-2013-1465-Yoshida, ps-91-2016-053002-Navratil,prc-100-2019-024304-Vorabbi}
and
rotation~\cite{plb-719-2013-179-Caprio,prc-91-2015-014310-Maris,*prc-99-2019-029902-Maris-ERRATUM,ijmpe-24-2015-1541002-Caprio,prc-93-2016-051301-Stroberg,prc-94-2016-011301-Jansen}
emerge from \textit{ab initio} calculations.
This leaves us with the question
of understanding the underlying physical nature of the collective correlations
giving rise to these patterns.

In a system exhibiting dynamical
symmetry~\cite{pr-125-1962-1067-gellmann,Wybourne1974,Barut1977,conmath-160-1994-151-Iachello,Iachello2015},
simple patterns arise naturally, since the spectrum of eigenstates is organized
according to irreducible representations (\textit{irreps}) of the dynamical
symmetry group.  In heavier nuclei, dynamical symmetries have played a central
role in characterizing nuclear correlations and collective
phenomena~\cite{prl-35-1975-1069-Arima,apny-99-1976-253-Arima,*apny-111-1978-201-Arima, *apny-115-1978-325-Scholten,*anpy-123-1979-468-Arima}.
In intermediate-mass nuclei, described by the shell model, Elliott's $\grpsu{3}$
dynamical
symmetry~\cite{prsla-245-1958-128-Elliott,*prsla-245-1958-562-Elliott,*prsla-272-1963-557-Elliott,*prsla-302-1968-509-Elliott,anp-1-67-1968-Harvey,arns-23-1973-123-Hecht}
provides a mechanism for the emergence of rotation.  In the lightest nuclei,
accessible by \textit{ab initio} theory, we may now seek to identify the role of
$\grpsptr\supset\grpu{3}$ dynamical
symmetry~\cite{prl-38-1977-10-Rosensteel,ap-126-1980-343-Rosensteel,rpp-48-1985-1419-Rowe}
in defining the structure of the excitation
spectrum.

The symplectic group $\grpsptr$, associated with the coordinates and momenta in
three dimensions, has long been proposed as an organizing scheme for the nuclear
many-body
problem~\cite{prl-38-1977-10-Rosensteel,ap-126-1980-343-Rosensteel,rpp-48-1985-1419-Rowe,ppnp-37-1996-265-Rowe}.
Through its $\grpu{3}$ subgroup, the symmetry group of the harmonic oscillator,
it is intimately connected to the nuclear shell model~\cite{prl-38-1977-10-Rosensteel,ap-126-1980-343-Rosensteel,plb-119-1982-249-Carvalho,npa-414-1984-93-Park,npa-413-1984-215-Draayer,npa-419-1984-1-Rosensteel,jmp-39-1998-5123-Escher,prl-84-2000-1866-Escher,prc-65-2002-054309-Escher}.
In its contraction limit, $\grpsptr$ yields a microscopic formulation of nuclear
collective dynamics, in terms of coupled rotational and giant monopole and
quadrupole vibrational degrees of
freedom~\cite{ps-91-2016-033003-Rowe,*ps-91-2016-049601-Rowe-ERRATUM,prc-101-2020-054301-Rowe}.

Wave functions obtained in \textit{ab initio} calculations have already been
identified as having specific dominant $\grpu{3}$ and $\grpsptr$ symmetry
components~\cite{prl-98-2007-162503-Dytrych,jpg-35-2008-095101-Dytrych,jpg-35-2008-123101-Dytrych,ppnp-67-2012-516-Draayer,prl-111-2013-252501-Dytrych,prc-91-2015-024326-Dytrych,cpc-207-2016-202-Dytrych,ppnp-89-2016-101-Launey,prl-124-2020-042501-Dytrych}.  
In this letter, calculations carried out in a symplectic no-core configuration
interaction (SpNCCI) framework demonstrate that the symmetry of individual
states moreover fits into an overall $\grpsptr\supset\grpu{3}$ dynamical
symmetry pattern of the spectrum as a whole.
 
In particular, for $\isotope[7]{Be}$, beyond the well-known $K=1/2$ ground state
rotational
band~\cite{plb-719-2013-179-Caprio,prc-91-2015-014310-Maris,prc-99-2019-029902-Maris-ERRATUM,ijmpe-24-2015-1541002-Caprio},
we find that an excited band emerges in the \textit{ab initio} calculations as a
symplectic collective excitation.  The remainder of the low-lying spectrum
follows an Elliott $\grpsu{3}$ dynamical symmetry pattern, where the rotational
structure is, however, dressed by multishell symplectic excitations.
Preliminary results were presented in
Refs.~\cite{aarsscps-3-2018-17-McCoy,McCoy2018}.

\paragraph{$\grpsptr\supset\grpu{3}$ dynamical symmetry.}

To recognize the role of $\grpsptr\supset\grpu{3}$ dynamical symmetry in
\textit{ab initio} calculated spectra, we must first be familiar with some 
basic properties of $\grpsptr$ irreps.  Elliott's
$\grpu{3}=\grpu{1}\times\grpsu{3}$ group considered here is the product of an
$\grpsu{3}$ generated by the orbital angular momentum
operator and a quadrupole tensor $\calQ$, and the $\grpu{1}$ group of the
harmonic oscillator Hamiltonian.  Then $\grpsptr$ augments
these generators with symplectic raising and lowering operators, which
physically represent creation and annihilation operators for giant monopole and
quadrupole vibrations.

\begin{figure*}
\begin{center}
\includegraphics[width=\ifproofpre{0.95}{1.0}\hsize]{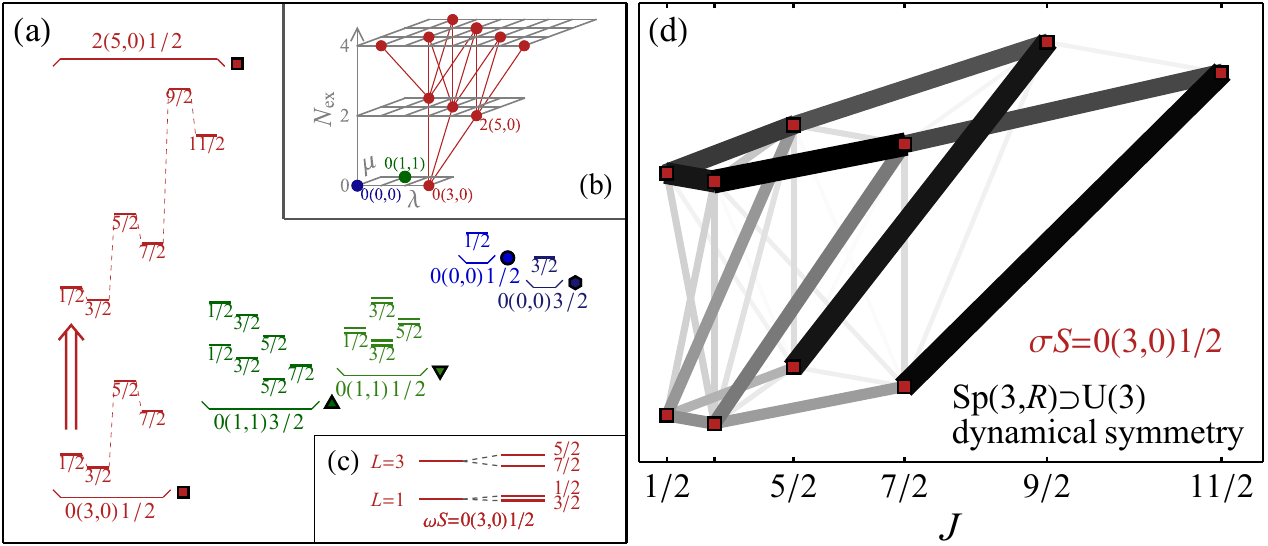}
\end{center}
~\\[-1.5\baselineskip]  
\caption{
Low-lying spectrum in an $\grpsptr\supset\grpu{3}$ dynamical symmetry description of 
  $\isotope[7]{Be}$.  (a)~Energies. Parameters in the dynamical symmetry
Hamiltonian~(\ref{eqn:H-dynamical}) are chosen for approximate consistency with
the experimental~\cite{npa-708-2002-3-Tilley} and \textit{ab initio} calculated
spectra of $\isotope[7]{Be}$.  States are grouped by $\grpu{3}$ irreps
(labeled by $\omega S$).  The excited $\omega S=2(5,0)1/2$ $\grpu{3}$ irrep 
obtained by symplectic raising (arrow) within the $\sigma S=0(3,0)1/2$ $\grpsptr$ irrep
is shown.  To facilitate comparison with
Fig.~\ref{fig:network-families}, $\grpsptr$ irreps are tagged by symbols
defined there for $\sigma S$.
(b)~Organization of $\grpsptr$ irrep $\sigma=0(3,0)$ into $\grpu{3}$
irreps~(dots), connected by the symplectic raising operator~(lines).  Shown
through $\Nex=4$.
(c)~Branching of the $\grpu{3}$ irrep $0(3,0)$ to orbital angular momenta $L$,
followed by coupling with spin ($S=1/2$) to give total angular momenta $J$.
(d)~Quadrupole transition strengths (isoscalar), within the $\sigma S=0(3,0)1/2$ $\grpsptr$ irrep,
with $B(E2)$ strength indicated by line thickness (and shading).
\label{fig:dynamical} 
}
\end{figure*}
        
An $\grpsptr$ irrep is comprised of an infinite tower of $\grpu{3}$ irreps.
Starting from a single $\grpu{3}$ irrep with some lowest number of oscillator
quanta, or \textit{lowest grade irrep} (LGI), the remaining $\grpu{3}$ irreps
are obtained by repeatedly acting with the symplectic raising operator, which
adds two oscillator quanta at a time.  Each $\grpu{3}$ irrep is characterized by
a fixed number of oscillator quanta and by $\grpsu{3}$ quantum numbers
$(\lambda,\mu)$, which are related to the nuclear
deformation~\cite{zpa-329-1988-33-Castanos}.  A $\grpu{3}$ irrep may therefore
be labeled by quantum numbers $\omega\equiv\Nwex(\lambda_\omega,\mu_\omega)$,
where $\Nex$ denotes the number of oscillator excitations relative to the lowest
Pauli-allowed oscillator configuration.  The entire $\grpsptr$ irrep is then
uniquely labeled by the $\grpu{3}$ quantum numbers
$\sigma\equiv\Nsex(\lambda_\sigma,\mu_\sigma)$ of its LGI.  The $2\hw$ and $4\hw$
$\grpu{3}$ irreps arising through the action of the symplectic raising operator
on the $\sigma=0(3,0)$ LGI in $\isotope[7]{Be}$ is illustrated in
Fig.~\ref{fig:dynamical}(b).

The full subgroup chain,
taking into account angular momenta, is
\begin{multline}
\label{eqn:sp-chain}
\bigl[
  \underset{\smash{\mathclap{\underset{\overbrace{\scriptstyle{\Nsex(\lambda_\sigma,\mu_\sigma)}}}{\sigma}}}}{\,\,\grpsptr\,\,\,\,}  
  \supset
  \underset{\smash{\mathclap{\underset{\overbrace{\scriptstyle{\Nwex(\lambda_\omega,\mu_\omega)}}}{\omega}}}}{\,\,\,\,\grpu{3}\,\,}  
  \supset
  \underset{L}{\grpso{3}}
\bigr]
\times
\underset{S}{\grpsu[S]{2}}
\supset\underset{J}{\grpsu[J]{2}}.
\end{multline}
Each $\grpu{3}$ irrep contains states of orbital angular momenta $L$ according
to the $\grpsu{3}\supset\grpso{3}$ branching rule~\cite{prsla-245-1958-128-Elliott}.
Fermionic antisymmetry defines the possible total spins
$S$~\cite{Hamermesh1962,anp-1-67-1968-Harvey,cpc-56-1989-279-Draayer} for each
$\grpu{3}$ irrep realized in the nuclear many-body system.  Then $L$ and $S$
combine to give total angular momenta $J$, as illustrated in
Fig.~\ref{fig:dynamical}(c) for the $\omega=0(3,0)$ irrep in
$\isotope[7]{Be}$, where $L=1,3$ combine with $S=1/2$ to give
$J=1/2,3/2,5/2,7/2$.

The low-energy spectrum expected in a dynamical symmetry description of
$\isotope[7]{Be}$ is illustrated in Fig.~\ref{fig:dynamical}(a).  In the $0\hw$
(or valence) space, the $\grpu{3}$ irreps which arise have $\omega=0(3,0)$,
$0(1,1)$ and $0(0,0)$, appearing in combination with specific spins $S$ as shown
in Fig.~\ref{fig:dynamical}(a).  Each serves as the LGI of an $\grpsptr$ irrep
(with $\Nsex=0$).  The $\grpu{3}$ irrep with $\omega S=2(5,0)1/2$ obtained by
symplectic laddering from the $\sigma S = 0(3,0)1/2$ LGI is shown.  Then, further $\grpsptr$
irreps originate at higher $\Nsex$.

The energy spectrum in Fig.~\ref{fig:dynamical} is determined by a
 simple dynamical symmetry Hamiltonian constructed from the Casimir operators
 for the subgroup chain~(\ref{eqn:sp-chain}):
\begin{equation}
\label{eqn:H-dynamical}
H=
\alpha C_{\grpsptr}
+ \varepsilon H_0 
+ \beta C_{\grpsu{3}}
+ a_L \Lvec^2 + a_S \Svec^2 + \xi \Lvec\cdot\Svec.
\end{equation}
Here, $C_{\grpsptr}$ is the Casimir operator of
$\grpsptr$~\cite{prc-65-2002-054309-Escher}, while the $\grpsu{3}$ Casimir
operator $C_{\grpsu{3}}=(1/6)(\calQ\cdot\calQ+3\Lvec^2)$ incorporates the
classic Elliott quadrupole Hamiltonian~\cite{anp-1-67-1968-Harvey}.  The
$\Jvec^2$ angular momentum Casimir operator is absorbed into
$\Lvec\cdot\Svec=\tfrac12(\Jvec^2-\Lvec^2-\Svec^2)$.

A $K=1/2$ ground-state band is experimentally observed in $\isotope[7]{Be}$ (and
mirror nuclide $\isotope[7]{Li}$)~\cite{rmp-25-1953-390-Inglis,npa-708-2002-3-Tilley},
with an exaggerated Coriolis energy staggering leading to an inverted angular
momentum sequence ($J=3/2,1/2,7/2,5/2$).  When the usual attractive sign is
taken on the quadrupole interaction in~(\ref{eqn:H-dynamical}), \textit{i.e.},
$\beta<0$, the leading (lowest energy) $\grpu{3}$ irrep is $0(3,0)$, which
indeed has the same angular momentum content [Fig.~\ref{fig:dynamical}(c)] as
the $\isotope[7]{Be}$ ground-state band.  The staggering is qualitatively
reproduced via the $\Lvec\cdot\Svec$ dependence in~(\ref{eqn:H-dynamical}).

Dynamical symmetry provides concrete predictions also for transition
strengths~\cite{conmath-160-1994-151-Iachello,Iachello2015}.  The isoscalar part 
of the quadrupole operator is a linear combination of $\grpsptr$ generators.
Thus, $\grpsptr\supset\grpu{3}$ dynamical symmetry implies strong $E2$
transitions between $\grpu{3}$ irreps differing by two quanta within an
$\grpsptr$ irrep.  Predictions for isoscalar $E2$ strengths follow directly from
$\grpsptr$ generator matrix
elements~\cite{jmp-21-1980-924-Rosensteel,jpa-17-1984-399-Rowe}, with no free
parameters, as illustrated in Fig.~\ref{fig:dynamical}(d) for $\sigma S=0(3,0)
1/2$.

\begin{figure*}
\begin{center}
\includegraphics[width=\ifproofpre{0.95}{1.0}\hsize]{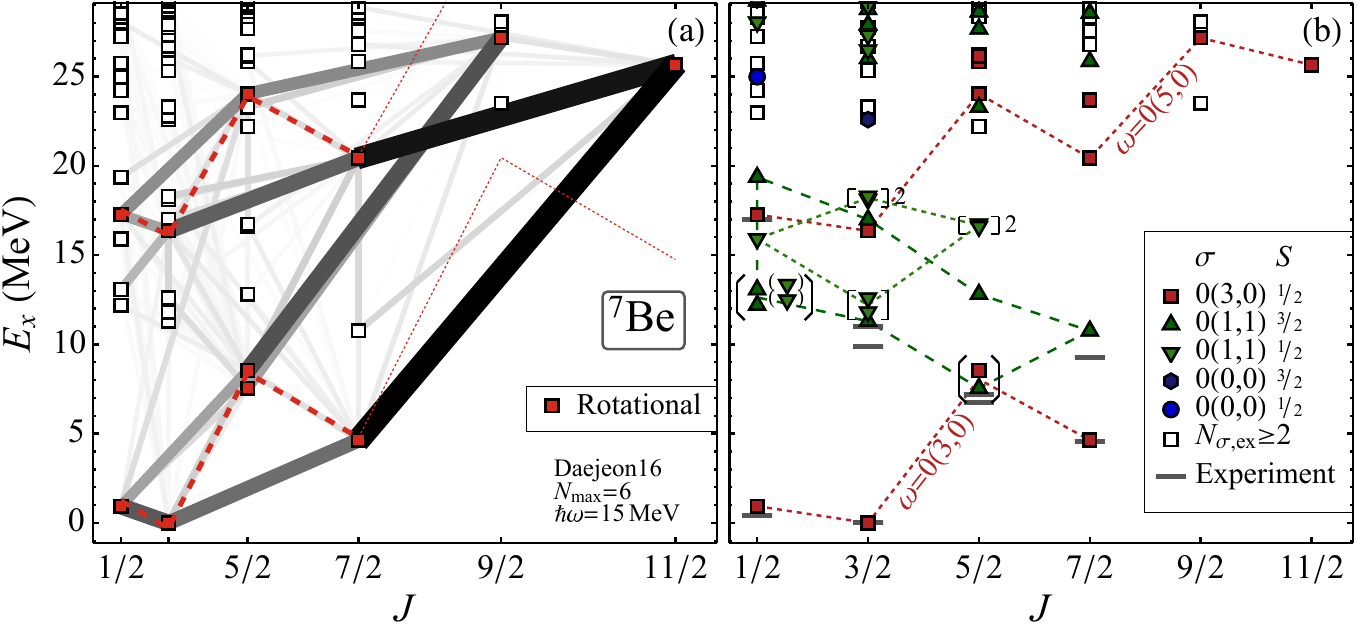}
\end{center}
~\\[-2\baselineskip]  
\caption{\textit{Ab initio} calculated negative parity energy spectrum of
  $\isotope[7]{Be}$: (a)~Rotational bands (red squares).  Strengths (line
  thickness and shading) are indicated for all $J$-decreasing $E2$ transitions
  from rotational band members (specifically,
  $J_f<J_i$ or $J_f=J_i$ and $E_f<E_i$).  Energies are plotted against angular momentum
  scaled as $J(J+1)$, as appropriate for rotational analysis. Fits to the  rotational
  energy formula with Coriolis staggering are
  shown (dashed/dotted lines).  (b)~Most significant $\grpsptr$ contribution $\sigma S$
  (indicated by symbol shape and color, see legend) for each state.  States with
  the same largest $\grpu{3}$ contribution $\omega S$ are connected by dashed
  lines.  Close-lying states may represent degenerate subspaces involving
  different internal spin couplings (square brackets, with a numeral 2
  indicating degenerate doublets indistinguishable in the plot) or may undergo
  significant two-state mixing (angled brackets).  Experimental
  energies~\cite{npa-708-2002-3-Tilley} are shown for context (horizontal lines).
  Calculation is for the Daejeon16 interaction, with $\Nmax=6$ and oscillator
  basis parameter $\hw=15\,\MeV$~\cite{Suhonen2007}.
\label{fig:network-families}
}
\end{figure*}
\begin{figure}
\begin{center}
\includegraphics[width=\ifproofpre{1}{0.5}\hsize]{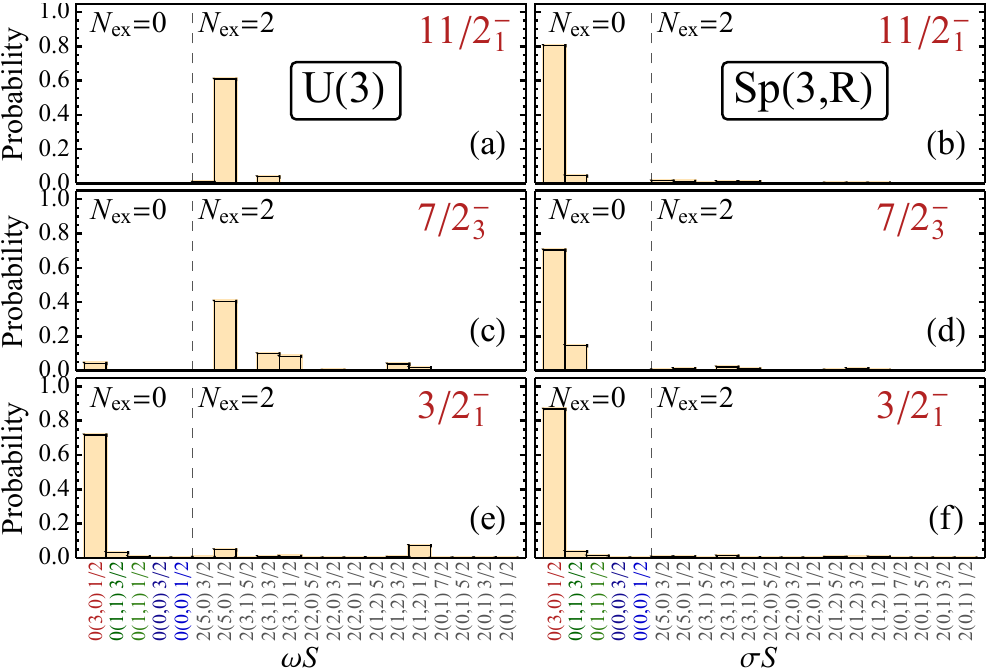}
\end{center}
~\\[-1.5\baselineskip]  
\caption{Decompositions of calculated $\isotope[7]{Be}$ wave functions by
  $\grpu{3}$~(left) and $\grpsptr$~(right) contributions, for the $\omega S=0(3,0)1/2$ ground state band member $3/2^-_1$~(bottom),
  the $\omega S=2(5,0)1/2$ excited band member $7/2^-_3$ (middle), and the strongly connected $11/2^-_1$~(top).  Contributions are arranged by $\Nex(\lambda,\mu)S$ and shown through
  $\Nex=2$.
\label{fig:decomposition}
}
\end{figure}

\paragraph{\textit{Ab initio} SpNCCI results for $\isotope[7]{Be}$.}

The present SpNCCI framework for~\textit{ab initio} calculations makes use of a
symmetry-adapted basis for the fermionic many-body space, one which reduces the
subgroup chain~\eqref{eqn:sp-chain} and is free of center-of-mass excitations.
Matrix elements of the Hamiltonian and other operators are obtained recursively
in terms of matrix elements between the LGIs, building on ideas of Reske,
Suzuki, and
Hecht~\cite{Reske1984,npa-455-1986-315-Suzuki,npa-448-1986-395-Suzuki}.  These
seed matrix elements are calculated using the $\grpu{3}$-coupled
symmetry-adapted no-core shell model
(SA-NCSM)~\cite{prl-111-2013-252501-Dytrych,cpc-207-2016-202-Dytrych}.  Details
may be found in Ref.~\cite{McCoy2018}.

Here we carry out SpNCCI calculations for $\isotope[7]{Be}$ with the Daejeon16
internucleon interaction~\cite{plb-761-2016-87-Shirokov}, in a basis
incorporating all $\grpsptr$ irreps with LGIs with up to $6$ quanta
($\Nsex\leq6$), and carrying each of these up to $6$ quanta ($\Nwex\leq6$), both
taken relative to the lowest Pauli-allowed configuration.  The resulting space
is simply the center-of-mass free
subspace~\cite{npa-897-2013-109-Luo,caprio2020:intrinsic} of the $\Nmax=6$ no-core
shell model (NCSM) space~\cite{ppnp-69-2013-131-Barrett}, and the
spectroscopic results, shown in Fig.~\ref{fig:network-families}(a), are
identical to those of a traditional $\Nmax=6$ calculation.

Although symmetry-adapted bases combined with physically motivated truncation
schemes can yield improved convergence of
calculations~\cite{jpg-35-2008-123101-Dytrych,cpc-207-2016-202-Dytrych}, our
interest here lies in understanding how the dynamical symmetry structure of
$\isotope[7]{Be}$ underlies the features of the
\textit{ab initio} calculated spectrum.  Since the basis reduces the
subgroup chain~\eqref{eqn:sp-chain},
SpNCCI calculations provide immediate access to the $\grpsptr$ and
$\grpu{3}$ symmetry decompositions of the calculated wave functions, as
illustrated in Fig.~\ref{fig:decomposition}. Further decompositions are provided
in the Supplemental Material~\cite{fn:supplement}.

Notably, rotational features emerge in the spectrum.  A $K=1/2$ ground state
band ($J=1/2$ through $7/2$) is readily recognized through enhanced $E2$
transitions in the \textit{ab initio} calculated spectrum
[Fig.~\ref{fig:network-families}(a), lower dashed line], as in earlier NCSM
calculations~\cite{plb-719-2013-179-Caprio,prc-91-2015-014310-Maris,prc-99-2019-029902-Maris-ERRATUM,ijmpe-24-2015-1541002-Caprio}.
Calculated excitation energies within the band are already largely insensitive
to $\Nmax$ even though absolute energies are not well-converged (see
Ref.~\cite{epja-56-2020-120-Caprio}).

Moreover, two higher angular momentum states ($9/2^-_2$ and $11/2^-_1$) have
strong $E2$ connections to this ground state band.  In previous NCSM
calculations~\cite{plb-719-2013-179-Caprio,prc-91-2015-014310-Maris,prc-99-2019-029902-Maris-ERRATUM,ijmpe-24-2015-1541002-Caprio},
these states have been considered as possible ground state band members, albeit
with energies above those expected from the standard rotational energy formula
with Coriolis staggering [Fig.~\ref{fig:network-families}(a), lower dotted
  line].  Their quadrupole moments are also anomalously large compared to the
ground state band members (Fig.~5 of Ref.~\cite{prc-91-2015-014310-Maris}).

However, these $9/2^-$ and $11/2^-$ states also have enhanced
transitions to a particular high-lying $5/2^-$ state and $7/2^-$ state, well off
the yrast line.  Tracing $E2$ strengths to lower $J$ reveals that these $5/2^-$
and $7/2^-$ states belong to an excited $K=1/2$ rotational band
[Fig.~\ref{fig:network-families}(a), upper dashed line].  One might therefore
suspect the $9/2^-$ and $11/2^-$ states belong to the excited rotational
band, albeit with energies below those expected for this
band [Fig.~\ref{fig:network-families}(a), upper dotted line].
  
Returning to the $\grpsptr\supset\grpu{3}$ decompositions of
Fig.~\ref{fig:decomposition} for insight, the wave functions of the ground state
band members are dominated by a single $\grpu{3}$ irrep, namely, $\omega
S=0(3,0)1/2$, as expected (above) from a dynamical symmetry picture.
About $60$--$70\%$ of the probability (or norm) of these states comes from this
$\grpu{3}$ irrep, as illustrated for the ground state
[Fig.~\ref{fig:decomposition}(e)], with the exception of the $5/2^-$ band
member, which lies in a close doublet and undergoes two-state mixing.

This $0\hw$ Elliott $\grpu{3}$ description of the ground state band is dressed
by $2\hw$ and higher excitations.  We see that excitations within the same
$\grpsptr$ irrep account for much of the remaining probability.  For the ground
state [Fig.~\ref{fig:decomposition}(f)], the $\sigma S =0(3,0)1/2$ $\grpsptr$
irrep accounts for over $80\%$ of the probability, which comes from,
\textit{e.g.}, the $\omega S=2(5,0)1/2$ and $2(1,2)1/2$ irreps within this
$\grpsptr$ irrep [recall Fig.~\ref{fig:dynamical}(b)].

For the excited band, the largest $\grpu{3}$ contribution comes from $\omega
S=2(5,0)1/2$, \textit{e.g.}, $\sim40\%$ for the $7/2^-$ band member
[Fig.~\ref{fig:decomposition}(c)]. This again suggests an Elliott rotational
description, but now in the $2\hw$ space rather than in the traditional $0\hw$
shell model valence space.  The $\grpu{3}$ symmetry is more diluted than
for the ground state band, and dressing with higher excitations is again
significant.

Moreover, we see that the excited band members lie almost entirely within the
same $\sigma S =0(3,0)1/2$ $\grpsptr$ irrep as the ground state band,
\textit{e.g.}, $\sim70\%$ for the $7/2^-$ band member
[Fig.~\ref{fig:decomposition}(d)].  While there are $8$ different
$\grpu{3}$ irreps with quantum numbers $\omega S=2(5,0)1/2$ for
$\isotope[7]{Be}$, the $2(5,0)1/2$ probability found in the calculated
wave function arises almost entirely from the one such $\grpu{3}$ irrep
lying in the $\sigma S =0(3,0)1/2$ symplectic irrep.

Thus, the wave functions are consistent with an approximate
$\grpsptr\supset\grpu{3}$ dynamical symmetry (Fig.~\ref{fig:dynamical}).  Indeed
the $\grpsptr$ symmetry is significantly better preserved than the $\grpu{3}$
symmetry.

Turning to the $9/2^-$ and $11/2^-$ states with strong
transitions to both bands, these have predominantly $\omega S=2(5,0)1/2$
$\grpu{3}$ content [Fig.~\ref{fig:decomposition}(a)], like the excited band
members but purer ($\sim50$--$60\%$).  They likewise lie almost entirely within the ground state's
$\sigma S =0(3,0)1/2$ $\grpsptr$ irrep [Fig.~\ref{fig:decomposition}(b)].

Thus, $\grpsptr\supset\grpu{3}$ dynamical symmetry provides a context for
understanding both the emergent rotational features and the incomplete
description provided for these features by a simple adiabatic rotational
picture.  Qualitatively, a $0\hw$ ground state band [$\omega S=0(3,0)1/2$] and
$2\hw$ excited band [$\omega S=2(5,0)1/2$] lie within the same symplectic irrep
[$\sigma S=0(3,0)1/2$].  In a pure Elliott $\grpu{3}$ rotational description,
the $9/2^-$ and $11/2^-$ states would simply be part of the excited band.
Enhanced transitions among these states are a consequence of dynamical symmetry
[Fig.~\ref{fig:dynamical}(d)], reflecting the role of the isoscalar $E2$
operator as a generator connecting states with $\Delta N=\pm 2$ within an
$\grpsptr$ irrep.

Yet, mixing of $\grpu{3}$ irreps within the $\grpsptr$ irrep, which becomes
significant for the off-yrast excited band members ($J\leq7/2$), manifests in
deviations from a pure Elliott rotational picture.  This breakdown is reflected
in weaker in-band and inter-band $E2$ transitions involving the low-$J$
excited band members [Fig.~\ref{fig:network-families}(a)], compared to the
dynamical symmetry predictions [Fig.~\ref{fig:dynamical}(d)], as well as the
discontinuity in energies between the low-$J$ and high-$J$ members of this band.

For the remaining low-lying states in Fig.~\ref{fig:network-families}(a),
the overall pattern of the spectrum is again qualitatively described by
$\grpsptr\supset\grpu{3}$ dynamical symmetry.  In
Fig.~\ref{fig:network-families}(b), the symbols identify the largest
$\grpsptr$ component, while dashed lines (where practical) connect states
sharing the same largest $\grpu{3}$ component.

For many of the states near the
yrast line the largest $\grpu{3}$ and $\grpsptr$ components contribute the
preponderance of the probability.  However, as we move to higher energy and away
from the yrast line, contributions from other $\grpu{3}$ and $\grpsptr$
components become increasingly important (see Supplemental
Material~\cite{fn:supplement}).
Furthermore, recall that, when two states are nearly degenerate
in energy, they may undergo two-state mixing [brackets in Fig.~\ref{fig:network-families}(b)].  This serves both to mix the
dynamical symmetry content and lift the energy degeneracy through level
repulsion~\cite{Casten2000,Heyde2020}.

The dynamical symmetry picture accounts for the full set of states in the
calculated low-lying ($0\hw$) spectrum and the overall pattern of their
energies.  In comparing Fig.~\ref{fig:dynamical}(a) with
Fig.~\ref{fig:network-families}(b), it is helpful to focus on the
``constellations'' formed when the calculated states are classified by their
predominant $\grpsptr\supset\grpu{3}$ contributions.  For instance, the
calculated states [Fig.~\ref{fig:network-families}(b)] with largest component
$\omega S=0(1,1)3/2$ form a roughly trapezoidal constellation (up triangles),
while those with $\omega S=0(1,1)1/2$ form two nearly degenerate diamond-shaped
constellations (down triangles), as in the dynamical symmetry picture
[Fig.~\ref{fig:dynamical}(a)].  Counterparts to the expected higher-lying
$\omega S=0(0,0)3/2$ (hexagon) and $\omega S=0(0,0)1/2$ (circle) are also found.

The relationship between the calculated spectrum and the dynamical symmetry
picture, shown here for the Daejeon16 interaction, is robust across choice of
internucleon interaction.  This is illustrated for, \textit{e.g.}, the
JISP16~\cite{plb-644-2007-33-Shirokov} and Entem-Machleidt \nthreelo{}
chiral perturbation theory~\cite{prc-68-2003-041001-Entem} interactions in the
Supplemental Material~\cite{fn:supplement}.

\paragraph{Conclusion.}

We have seen that $\grpsptr\supset\grpu{3}$ dynamical symmetry, as laid out in
Fig.~\ref{fig:dynamical}, provides an organizing scheme for understanding the
entire low-lying \textit{ab initio} calculated spectrum of $\isotope[7]{Be}$,
as shown in Fig.~\ref{fig:network-families}.  Symmetry is reflected not
merely in the decompositions of the individual wave functions
(Fig.~\ref{fig:decomposition}), but in the overall arrangement of energies,
which is remarkably consistent with a simple dynamical symmetry
Hamiltonian~(\ref{eqn:H-dynamical}), and in the $E2$ transition patterns among
them.

Essential features of the dynamical symmetry structure are: a $0\hw$ spectrum
well-described in an Elliott $\grpu{3}$ picture, but dressed by $2\hw$ and
higher contributions from within the $\grpsptr$ irrep; and excited states
reflecting $2\hw$ excitations within the $\grpsptr$ irrep, related to the
lower-lying states by strong quadrupole transitions.  Although the purity of
$\grpu{3}$ symmetry falls off away from the yrast line, as reflected in the
inability of a simple rotational description to simultaneously describe both the
low-$J$ members of the excited band and the strongly connected $9/2^-$ and
$11/2^-$ states, the persistence of $\grpsptr$ symmetry explains the presence of
these strong transitions.

The connection between the ground state and excited bands by the symplectic
raising operators, which physically represent creation operators for the giant
monopole and quadrupole resonances, is suggestive of the emergence of collective
vibrational degrees of freedom.  In the light, weakly bound $\isotope[7]{Be}$
system, such an interpretation can at most be approximate.  In the present bound
state formalism, it moreover remains uncertain how the structure of the
calculated excited band relates to the structure of physical
resonances~\cite{prc-100-2019-024304-Vorabbi}.

Nonetheless, the presence of rotational bands with strong $E2$ connections may
be taken as a possible precursor to rotational-vibrational structure in
heavier and more strongly bound systems.  Indeed, the emergence of $\grpsptr$
symmetry as an organizing scheme for nuclear structure in light nuclei provides
a link to more purely collective interpretations of the dynamics through the
rotation-vibration degrees of freedom which naturally arise in the classical
(large quantum number) limit of the symplectic
description~\cite{plb-140-1984-155-Le_Blanc,rpp-48-1985-1419-Rowe,npa-504-1989-76-Rowe,npa-452-1986-263-Le_Blanc,ps-91-2016-033003-Rowe}.



\section*{Acknowledgements}

\begin{acknowledgments}
We thank David J.~Rowe for indispensable assistance with the $\grpsptr$ formalism,
Calvin W.~Johnson, James P.~Vary, and Pieter Maris for valuable
discussions, and Nadya A.~Smirnova and Jakub Herko for comments on the manuscript.
This material is based upon work supported by the U.S.~Department of Energy,
Office of Science, Office of Nuclear Physics, under Award Number
DE-FG02-95ER-40934, by the U.S.~Department of Energy, Office of Science, Office
of Workforce Development for Teachers and Scientists, Graduate Student Research
(SCGSR) program, under Contract Number DE-AC05-06OR23100, and by the Research
Corporation for Science Advancement, under a Cottrell Scholar Award.  TRIUMF receives federal
funding via a contribution agreement with the National Research Council of
Canada.
This research used computational resources of the University of Notre Dame
Center for Research Computing and of the National Energy Research Scientific
Computing Center (NERSC), a U.S.~Department of Energy, Office of Science, user
facility supported under Contract~DE-AC02-05CH11231.
\end{acknowledgments}

%


\end{document}